\pdfoutput=1

\input{aipcheck}

\documentclass[
    ,final            
  ]
  {aipproc}

\layoutstyle{8x11single}

\usepackage{graphicx}

\begin{document}

\title{Prospects for X(3872) Detection at ${\sf \overline{P}ANDA}$}


\author{Jens S\"oren Lange, Martin Galuska, Thomas Ge\ss{}ler, Wolfgang K\"uhn,\\
Stephanie K\"unze, Yutie Liang, David M\"unchow, Bj\"orn Spruck,\\
Matthias Ullrich, and Marcel Werner}{
  address={Justus-Liebig-Universit\"at Gie\ss{}en, II.\ Physikalisches Institut, 
  Heinrich-Buff-Ring 16, 35392 Gie\ss{}en, Germany}
}

\begin{abstract}
Monte-Carlo simulations for a resonance scan of the charmonium-like state X(3872) 
at ${\sf \overline{P}ANDA}$ are performed.
Final state radiation hadronic background reactions are taken into account.
The signal reconstruction
uses a realistic pattern recognition (track finder and track fitter) 
and electron/pion discrimination.
\end{abstract}

\maketitle

\noindent
Conference Talk presented at MENU10, 12$^{th}$ International Conference on Meson-Nucleon Physics 
and the Structure of the Nucleon, 05/31-06/04, 2010, College of William and Mary, Williamsburg, Virginia.

\section{The X(3872) State}

\noindent
The X(3872) is a charmonium(-like) state which was 
first observed in its decay X(3872)$\rightarrow$J/$\psi$$\pi^+$$\pi^-$
in $B$ decays \cite{x3872_belle} \cite{x3872_babar} 
and inclusive production in $p$$\overline{p}$ collisions 
\cite{x3872_cdf} \cite{x3872_d0}.
Meanwhile more decays such as final states 
$D^{0}$$\overline{D}^{0*}$ or $J$/$\psi$$\omega$ have been found.
In particular, the observation of its radiative decay into $J$/$\psi$$\gamma$ 
allows the assignment of positive charge parity $C$=$+$1.
The interesting observation of isospin violation in the decay
X(3872)$\rightarrow$$J$/$\psi$$\rho$($\rightarrow$$\pi^+$$\pi^-$) 
raised the question, if it might be a charmonium state at all.
As its mass is within $\Delta$$m$$<$1~MeV of the $D^{0}$$\overline{D}^{0*}$ threshold, 
it might be interpreted as an $S$-wave molecular state \cite{tornqvist_1} \cite{tornqvist_2}.
Many properties of the X(3872) are being studied at ongoing experiments.
However, the width is unknown and probably can only be determined at ${\sf \overline{P}ANDA}$.
The current upper limit is $\Gamma$<2.3~MeV \cite{x3872_belle}, whereas
${\sf \overline{P}ANDA}$ might be able to set an upper limit
in the order of a few hundred keV. The quantitative 
investigation of this upper limit on the width  
by detailed Monte-Carlo (MC) simulations incl.\ background 
is topic of this paper.

\section{Charmonium Production at ${\sf \overline{P}ANDA}$}

\noindent
The ${\sf \overline{P}ANDA}$ 
(Anti-\underline{P}roton \underline{An}nihilation at \underline{Da}rmstadt)
experiment will be located at the future 
FAIR (Facility for Antiproton and Ion Research) facility \cite{fair}
at GSI in Darmstadt, Germany.
The primary synchrotron will have a circumference of 1.084~km, 
a proton beam with 30$\leq$$E_{beam}$$\leq$90~GeV 
and an intensity $\leq$2$\times$10$^{13}$/s.
Antiprotons are produced by a secondary target 
and then stored and cooled in the HESR (High Energy Storage Ring).
There are two HESR modes.
In the high intensity mode (high resolution mode), 
there are 10$^{11}$ (10$^{10}$) stored antiprotons, 
and stochastic cooling (electron cooling) will provide a 
beam momentum resolution of 
$\Delta$$p$/$p$$\geq$1$\cdot$10$^{-4}$ ($\Delta$$p$/$p$$\geq$4$\cdot$10$^{-5}$).
${\sf \overline{P}ANDA}$ will investigate $p$$\overline{p}$ collisions with an antiproton beam 
momentum $p_{beam}$$\leq$15 GeV/c, corresponding to $\sqrt{s}$$\leq$5.5~GeV.
In $e^+$$e^-$ collisions, only states with quantum numbers $J^{PC}$=$1^{--}$
can be produced directly due to the intermediate virtual photon.
As the X(3872) has positive $C$ parity, thus, it can not be produced directly
in $e^+$$e^-$.
However, in $p$$\overline{p}$ collisions two or three gluons are produced
in the annihilation, and states with any quantum number can be formed.

\noindent
At the ${\sf \overline{P}ANDA}$ interaction region, three different types of targets
are foreseen, namely 
{\it (a)} hydrogen pellets with a thickness of $\simeq$25~$\mu$m
and a falling velocity of $\geq$60~m/s, 
{\it (b)} a cluster jet target and
{\it (c)} fixed nuclear targets such as $Be$, $C$, $Si$ or $Al$.
For the pellet target, a peak luminosity of 
${\cal L}$=2$\times$10$^{32}$~cm$^{-2}$~s$^{-1}$ is expected.
If adjusted to a single resonance, such as e.g.\ the J/$\psi$
this would correspond to about 2$\times$10$^{9}$ produced J/$\psi$ per year.

\noindent
A resonance scan with the cooled beam could 
provide a measurement of the X(3872) width.
This technique was pioneered by the Fermilab experiments E760 and E835. 
With a beam momentum resolution of 
$\Delta$$p$/$p$=2$\cdot$10$^{-4}$, corresponding to a 
$\sqrt{s}$ resolution of 0.5~MeV (FWHM), 
the measurement of the width of the $J$/$\psi$ with 
$\Gamma$($J$/$\psi$)=99$\pm$12(stat.)$\pm$6(syst.)~keV
and the width of the $\psi'$ with 
$\Gamma$($\psi'$)=306$\pm$36(stat.)$\pm$16(syst.)~keV
could be performed \cite{e760}.

\noindent
The requirements for the ${\sf \overline{P}ANDA}$ experiment are quite high:
capability for high rate detection and data acquisition for 
$\leq$2$\times$10$^7$ interactions/s, 
$\simeq$4$\pi$ solid angle coverage (incl.\ a high resolution forward detector,
as ${\sf \overline{P}ANDA}$ is a fixed target experiment), 
secondary vertex detection of e.g.\ $D$ mesons with a resolution $\sigma$<100~$\mu$m, 
charged track momentum resolution $\Delta$p/p$\simeq$1\%,
charged particle identification for $e$$^{\pm}$, $\mu$$^{\pm}$, $\pi$$^{\pm}$, $K$$^{\pm}$, $p$$^{\pm}$
using e.g.\ Cherenkov detectors with internal reflection, 
and electromagnetic calorimetry in the range 10 MeV<$E_{\gamma}$<5~GeV.
Details about the experiment can be found elsewhere \cite{panda}.
For the down below results, 
MC simulations were performed with all ${\sf \overline{P}ANDA}$ subdetectors implemented,
but the reconstruction used only the following ${\sf \overline{P}ANDA}$ subdetectors:

\begin{itemize}

\item a Time Projection Chamber (TPC) with 135 padrows (corresponding 
to $\leq$135 hits for a charged particle track) and 135,169 pads of 2$\times$2~mm$^2$ area, 

\item a Micro Vertex Detector (MVD) with 120 silicon pixel modules 
(100$\times$100~$\mu$m$^2$ pixel size
and in total 10$^7$ readout channels), and 400 silicon strip modules  
($\simeq$0.5 m$^2$ active area and 7$\times$10$^4$ readout channels), and 

\item an Electromagnetic Calorimeter (EMC) with 
$\simeq$17,200 crystals of PbWO$_4$, a radiation hard scintillation material 
with a fast decay constant of $\simeq$6~ns. The thickness corresponds 
to $\simeq$28$X_0$ radiation lengths.

\end{itemize}

\noindent
The MC simulation was performed with the PandaRoot \cite{pandaroot}
simulation, digitization, reconstruction and analysis framework.
It consists of $\simeq$43,000 geometry volumes 
(incl. details such as e.g.\ the beampipe for the pellet target) 
and $\geq$400,000 lines of C++ code.
Transport engines are Geant3, Geant4, and Fluka. 
Event Generators are EvtGen, DPM, PYTHIA, and UrQMD. 
References can be found elsewhere \cite{pandaroot_chep08}.
PandaRoot is used on $\geq$15 Linux platforms.

\noindent
Recent improvements, in particular since 
the ${\sf \overline{P}ANDA}$ Physics Report \cite{panda_physics_book} are
in particular: 
{\it (a)} for the first time X(3872) simulations, 
not investigated before as a physics channel for ${\sf \overline{P}ANDA}$, 
{\it (b)} usage of detailed field maps 
of a homogenous $B_z$=2~T in the central region 
and dipole field of 2~Tm in the forward region 
(with flux effects in the iron of the ${\sf \overline{P}ANDA}$ muon subdetector
and interference of the fields taken into accout), 
{\it (c)} usage of a realistic track finder and track fitter 
based upon a conformal map technique, and
{\it (d)} simulation of final state radiation \cite{photos}.
In about $\simeq$30\% of all $J$/$\psi$$\rightarrow$$e^+$$e^-$ decays an additional
photon is radiated.

\noindent
The particle indentification uses the ratio $E$/$p$ 
as a variable for discrimination of charged pions and leptons.
$E$ is the deposited shower energy in the EMC,
$p$ is the reconstructed track momentum by the track finder
and track fitter using the MVD and the TPC.
For leptons $E$/$p$ is $\simeq$1.
For charged pions, which generate hadronic showers,
$E$/$p$ can take any value 0$\leq$$E$/$p$$\leq$1.
By applying an $E$/$p$ cut of 0.8$\leq$$E$/$p$$\leq$1.2 for electrons, 
the combinatorial background is reduced by $\geq$90\%.

\section{Estimated Rates for X(3872) Formation at ${\sf \overline{P}ANDA}$}

For resonant X(3872) formation, the required beam momentum
is $p_{beam}$=6.99100~GeV/c.
Our baseline assumption is a peak cross section of $\sigma_{X(3872)}$=50~nb,
in the same order of magnitude as e.g.\ the cross section of $\psi'$ production \cite{e760}.
Note, that, if the X(3872) is a molecular $D^0$$\overline{D}^{0*}$ state,
there are estimates \cite{chen_ma} that the cross section may increase up to $\leq$443~nb.
The ratio of branching fractions of the X(3872) decays 
into the final states $D^0$$\overline{D}^{0*}$:$J$/$\psi$$\pi^+$$\pi^-$:$J$/$\psi$$\gamma$ 
is assumed to be 9:1:0. 
All other decays are assumed to have a zero branching fraction.
The branching fraction for $J$/$\psi$$\rightarrow$$e^+$$e^-$ and 
$\mu^+$$\mu^-$ is $\simeq$6\% each.
The reconstruction efficiency of $\simeq$50\% is 
dominated by the track reconstruction efficiency 
for the low momentum charged pions. 
All these factors lead to a reconstructable cross section of 250~pb.
For resonance scans, the HESR will be operated in the 
high resolution mode with $\Delta$p/p=10$^{-5}$, which corresponds
to a luminosity of ${\cal L}$=2$\times$10$^{31}$~cm$^{-2}$s$^{-1}$.
Assuming an accelerator duty factor of 50\%, 
this leads to an integrated luminosity of 
${\cal L}_{int}$=0.86~pb$^{-1}$/day.
For a resonance scan with 20 energy points and 2 days/point,
this would correspond to a yield of $\simeq$215 events of 
$p$$\overline{p}$$\rightarrow$X(3872)$\rightarrow$$J$/$\psi$$\pi^+$$\pi^-$ 
per day at peak.

\section{Background}

\noindent
The main background is meson production 
in processes such as 
$p$$\overline{p}$$\rightarrow$$\pi^+$$\pi^-$$\pi^+$$\pi^-$
with two misidentified charged pions (as leptons) in the EMC.
The cross section for this process is 50~$\mu$b \cite{background}, 
compared to an estimated signal cross section of 50~nb.
For comparison, the total $p$$\overline{p}$ cross section
is $\simeq$70~mb.
It is important to investigate the shape of the background
e.g.\ in the 2-particle mass spectrum.
For this purpose, a DPM (dual parton model) event generator 
\cite{dpm_1} \cite{dpm_2} \cite{dpm_3} was used.
In fact, a varying background shape in the region of the $J$/$\psi$
was found and fitted with a first order Chebyshev polynome.
The fit was performed differently for each beam momentum in the resonance
scan.

\section{Determination of a Resonance Width by Resonance Scan}

The Breit-Wigner cross section for the formation and subsequent decay of a 
$c$$\overline{c}$ resonance $R$ of spin $J$, mass $M_R$ and total 
width $\Gamma_R$ formed in the reaction $\overline{p}$$p$$\rightarrow$$R$ is

\begin{equation}
\sigma_{BW} ( E_{cm} ) =
\frac{(2J+1)}{(2S+1)(2S+1)}
\frac{4 \pi (\hbar c)^2}{ ( E_{cm}^2 - 4 (m_p c^2)^2 )}
\times 
\frac{\Gamma_R^2 BR(\overline{p} p\rightarrow R) \times BR(R\rightarrow f)}
{(E_{cm} - M_R c^2 )^2 + \Gamma_R^2/4}
\end{equation}

where $S$ is the spin of the (anti-)proton.

\begin{equation}
\sigma ( E_{cm} ) =
\int_0^{\infty} 
\sigma_{BW} ( E' ) G (E' - E_{cm} ) dE'
\end{equation}

is a convolution of a Breit-Wigner term for the resonance
and the function $G$ for the beam resolution.

The area under the resonance peak is given by

\begin{equation}
A = \int_0^{\infty} 
\sigma ( E_{cm} ) dE_{cm} =
\frac{\pi}{2} \sigma_{peak} \Gamma_{R}
\end{equation}

which importantly is independent of the form of $G$($E$).
$\sigma_{peak}$ is the cross section at $E_{cm}$=$M_Rc^2$ given by

\begin{equation}
\sigma_{peak} =
\frac{(2J+1)}{(2S+1)(2S+1)}
\frac{16 \pi \hbar^2  
BR(\overline{p} p\rightarrow R) \times BR(R\rightarrow f) 
}{
( M_R - 4 m_p^2 ) c^2 \quad .
}
\end{equation}

By measuring $A$ using a fit to the excitation function 
and inserting $\sigma_{peak}$ into Eq.~3, the resonance width $\Gamma_R$
can be determined.

\section{Preliminary Results}

\begin{figure}[htb]
\unitlength1cm
\begin{picture}(17,12)
\centerline{\includegraphics[width=\textwidth,bb=0 0 607 442]{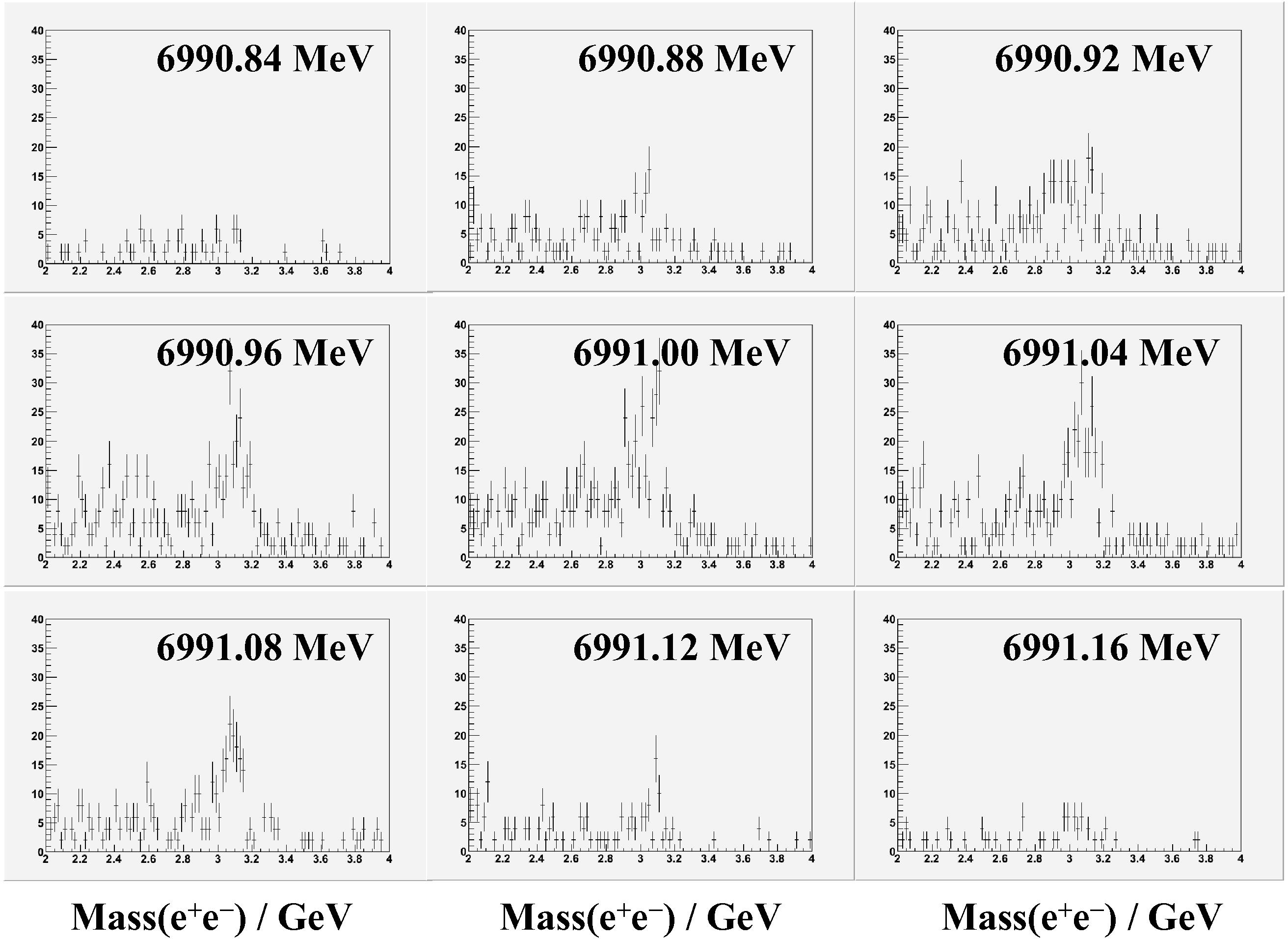}}
\end{picture}
\caption{MC simulation of a resonance scan of the X(3872) at ${\sf \overline{P}ANDA}$, 
with 9 scan points of different anti-proton beam momenta
and data taking of 2 days for each point.
The plots show the background subtracted, tagged $J$/$\psi$ signal from the 
X(3872)$\rightarrow$$J$/$\psi$$\pi^+$$\pi^-$ decay. Only statistical errors are shown.
}
\end{figure}

MC simulations for 
$p$$\overline{p}$$\rightarrow$X(3872)$\rightarrow$$J$/$\psi$$\pi^+$$\pi^-$
with background were performed 
for 9 different beam momenta in the 
resonance region.
Fig.~1 shows the $e^+$$e^-$ invariant mass with PID applied
and background subtracted. 
The final state is exclusive, i.e.\ no other particle is produced
except for photons from final state radiation 
(visible as radiative tail at masses smaller than the $J$/$\psi$ mass).
Each data point corresponds to 2 days of data taking.
A Gaussian fit was used 
to extract the signal yield. This $J$/$\psi$ yield is equal to the tagged X(3872) yield.
Then data points were fitted with an excitation function. 
The preliminary results are as follows:

\begin{itemize}

\item The fitted width of the excitation function
is $\simeq$20\% larger than the width which was used as input
in the MC simulation, due to the analysis technique such as  
the background subtraction and the reconstruction technique 
(e.g.\ the track fitter).
Note, that the beam resolution is not the limit for the 
width measurement. Even with a beam momentum resolution of $\Delta$$p_{beam}$$\simeq$0.5~MeV,
the E760 experiment was able to determine the factor of $\simeq$5 smaller width 
of the $J$/$\psi$ \cite{e760}. 
The difference between generated and reconstructed width
would only be a systematic error on the width measurement.

\item The width of the excitation function is still $\leq$100 keV, 
if the assumed resolution of the anti-proton beam 
is $\Delta$$p$/$p$=2$\cdot$10$^{-5}$
(HESR high resolution mode). 
However, in this analysis all distributions were assumed to have a Gaussian shape, 
while the correct resonance shape would be given by a Breit-Wigner parametrisation
and the shape of the function $G$($E$) in Eq.~2 might be non-Gaussian.
In addition, for our analysis, we approximated the function 
$\sqrt{s}$=$\sqrt{2 m_p^2 + 2 m_p \sqrt{p_{beam}^2 + m_p^2}}$
by a linear function in the small range of the resonance region.
As long as the beam resolution is accurately 
known, our results indicate that the width of the X(3872)
could be determined at ${\sf \overline{P}ANDA}$, 
if it is larger than $\Gamma_R$$\geq$100~keV and 
if the DPM generator describes 
the background shape correctly.

\end{itemize}

\noindent
These results are still preliminary, as the unfolding (integral equation, Eq.~2)
is ongoing work.


\bibliographystyle{aipproc}   
\bibliography{sample}

\IfFileExists{\jobname.bbl}{}
 {\typeout{}
  \typeout{******************************************}
  \typeout{** Please run "bibtex \jobname" to optain}
  \typeout{** the bibliography and then re-run LaTeX}
  \typeout{** twice to fix the references!}
  \typeout{******************************************}
  \typeout{}
 }

\end{document}